\documentclass[prd,nofootinbib,reprint,superscriptaddress,showkeys,showpacs]{revtex4-1}

\usepackage{amsmath}
\usepackage{amsfonts}
\usepackage{bbm}
\usepackage{hyperref}
\usepackage{cleveref}
\usepackage{color}
\usepackage{microtype}
\usepackage{graphicx}

\newcommand{\be}{\begin{equation}}
\newcommand{\ee}{\end{equation}}

\makeatother

\begin{document}

\title{Integrated Sachs-Wolfe effect in massive bigravity}
\author{Jonas Enander}
\email{enander@fysik.su.se}
\affiliation{The Oskar Klein Centre for Cosmoparticle Physics, Stockholm University\\
AlbaNova University Center, SE 106 91 Stockholm, Sweden}
\affiliation{Department of Physics, Stockholm University\\
AlbaNova University Center, SE 106 91 Stockholm, Sweden}
\author{Yashar Akrami}
\email{y.akrami@thphys.uni-heidelberg.de}
\affiliation{Institut f\"ur Theoretische Physik, Ruprecht-Karls-Universit\"at Heidelberg\\
Philosophenweg 16, 69120 Heidelberg, Germany}
\author{Edvard M\"ortsell}
\email{edvard@fysik.su.se}
\affiliation{The Oskar Klein Centre for Cosmoparticle Physics, Stockholm University\\
AlbaNova University Center, SE 106 91 Stockholm, Sweden}
\affiliation{Department of Physics, Stockholm University\\
AlbaNova University Center, SE 106 91 Stockholm, Sweden}
\author{Malin Renneby}
\email{malin.renneby@physik.uni-muenchen.de}
\affiliation{Excellence Cluster Universe, Ludwig-Maximilians-Universit\"at\\
Boltzmannstra{\ss}e 2, 85748 Garching, Germany}
\author{Adam R. Solomon}
\email{a.r.solomon@damtp.cam.ac.uk}
\affiliation{DAMTP, Centre for Mathematical Sciences, University of Cambridge\\
Wilberforce Rd., Cambridge CB3 0WA, UK}

\begin{abstract}

We study the integrated Sachs-Wolfe (ISW) effect in ghost-free, massive bigravity. We focus on the infinite-branch bigravity (IBB) model which exhibits viable cosmic expansion histories and stable linear perturbations, while the cosmological constant is set to zero and the late-time accelerated expansion of the Universe is due solely to the gravitational interaction terms. The ISW contribution to the CMB auto-correlation power spectrum is predicted, as well as the cross-correlation between the CMB temperature anisotropies and large-scale structure. We use ISW amplitudes as inferred from the WMAP 9-year temperature data together with galaxy and AGN data provided by the WISE mission in order to compare the theoretical predictions to the observations. The ISW amplitudes in IBB are found to be larger than the corresponding ones in the standard $\Lambda$CDM model by roughly a factor of 1.5, but are still consistent with the observations.

\end{abstract}



\maketitle

\section{Introduction}
\label{sec:intro}

The theoretical and phenomenological implications of massive gravity and its bimetric extension have been extensively studied since the ghost-free, nonlinear formulation was constructed in 2011~\cite{deRham:2010ik,deRham:2010kj,deRham:2011rn,deRham:2011qq,Hassan:2011vm,Hassan:2011hr,Hassan:2011tf,Hassan:2011zd,Hassan:2011ea}, some 70 years after the linear theory of massive gravity was proposed by Fierz and Pauli~\cite{Fierz:1939ix} (see Ref.~\cite{deRham:2014zqa} for a comprehensive review). It has been shown that in order for the nonlinear theory to avoid the so-called Boulware-Deser (BD) ghost~\cite{Boulware:1973my}, one needs to extend the gravitational sector by introducing at least one extra metric-like spin-2 tensor field. This so-called reference metric can either be fixed, resulting in a theory of massive gravity where gravitons possess five degrees of freedom, or be given dynamics, giving rise to a bimetric theory with two interacting tensor fields and seven degrees of freedom, corresponding to one massless and one massive graviton. The two theories are often referred to, respectively, as the de Rham-Gabadadze-Tolley theory of massive gravity, or dRGT, and the Hassan-Rosen theory of massive bigravity.

Soon after it was proposed, the original dRGT theory, where only the dynamical metric couples to matter, was shown to suffer from a no-go theorem forbidding flat and closed Friedmann-Lema\^itre-Robertson-Walker (FLRW) cosmological solutions on a flat reference metric~\cite{D'Amico:2011jj}. Solutions with an open FLRW metric or a non-flat reference metric do not save the theory, as they suffer from the so-called Higuchi instability~\cite{Higuchi:1986py} or other types of instabilities~\cite{Gumrukcuoglu:2011ew,Gumrukcuoglu:2011zh,Vakili:2012tm,DeFelice:2012mx,Fasiello:2012rw,DeFelice:2013awa}. Several solutions to these problems have been proposed in the literature by either adding extra degrees of freedom to dRGT~\cite{D'Amico:2012zv,D'Amico:2011jj,Huang:2012pe,Jaccard:2013gla,Foffa:2013vma,Dirian:2014ara,Comelli:2013paa,Comelli:2013tja} (for example stable and self-accelerating cosmological solutions are possible in the quasidilaton extension of massive gravity~\cite{Gabadadze:2014kaa,Gabadadze:2014gba}) or assuming inhomogenous and/or anisotropic metrics which are arbitrarily close to FLRW solutions on observable scales~\cite{D'Amico:2011jj,Volkov:2012cf,Volkov:2012zb,Gratia:2012wt,deRham:2014zqa}. As a different route to overcome the no-go theorem, it has recently been shown in Refs.~\cite{deRham:2014naa,deRham:2014fha,Gumrukcuoglu:2014xba} that exact FLRW solutions in dRGT can exist if at least some matter couples to both the dynamical and the fixed reference metrics through a generalized matter coupling (see, however, Ref.~\cite{Solomon:2014iwa} for complications this approach would introduce for cosmological studies). The hybrid metric introduced in this extended theory does, however, reintroduce the BD ghost above a certain energy scale,\footnote{This scale is, however, thought to be above the strong coupling scale of the theory. Therefore phenomenology of the theory, including its cosmological implications, can likely be safely studied below a cut-off scale larger than the strong coupling scale~\cite{deRham:2014naa}.} which is a generic feature of theories where both metrics couple to matter simultaneously~\cite{Yamashita:2014fga,deRham:2014naa,Noller:2014sta,Hassan:2014gta,deRham:2014fha,Soloviev:2014eea,Heisenberg:2014rka}, including the simple case of coupling the two metrics to matter minimally~\cite{Akrami:2013ffa,Akrami:2014lja,Khosravi:2011zi,Tamanini:2013xia}.

In this paper we consider the Hassan-Rosen theory of massive bigravity in its original formulation, i.e., where both metrics are dynamical while only one of them couples to matter. Many of the problems mentioned above are cured in bigravity, making it an attractive modification to general relativity (GR) with interesting phenomenology. The cosmology of bigravity has been extensively studied at both the background and perturbative levels (see, e.g., Refs.~\cite{vonStrauss:2011mq,Comelli:2012db,Khosravi:2012rk,Berg:2012kn,Akrami:2012vf,Akrami:2013pna,Enander:2013kza,Fasiello:2013woa,Konnig:2013gxa,Konnig:2014dna,Comelli:2014bqa,DeFelice:2014nja,Solomon:2014dua,Konnig:2014xva,Lagos:2014lca,Cusin:2014psa}), and the theory has been shown to be consistent with observed lensing and dynamical properties of local sources such as elliptical galaxies~\cite{Enander:2013kza}. In particular, it has been shown that the theory can provide viable cosmological models as alternatives to the standard $\Lambda$CDM model, consistent with all existing cosmological observations while offering clear signatures that can be verified using future large-scale structure experiments (see also Refs.~\cite{Enander:2014xga,Schmidt-May:2014xla,Comelli:2015pua} for cosmological studies of bigravity in the context of the generalized matter coupling). Perturbative studies have placed strong conditions on the parameter space of the theory as a large fraction of the models generate early-time instabilities which, although they do not necessarily rule these models out, complicate their comparison to observations as linear perturbation theory cannot be employed. A specific submodel, infinite-branch bigravity (IBB)~\cite{Konnig:2014xva}, exhibits both viable background solutions and stable linear perturbations that are consistent with observational data. This is the model that we focus on in this paper.\footnote{During the completion of this paper, an update to Ref.~\cite{Lagos:2014lca} appeared which studied the Higuchi bound in IBB and showed that it is not satisfied (see also Ref.~\cite{Konnig:2015lfa} which appeared later). This can potentially be dangerous since instabilities might appear at higher-order perturbations. Whether that is the case or not needs to be studied by analyzing nonlinearities; we leave the investigation of this for future work.}

As a continuation of the cosmological studies of bigravity, here we perform an analysis of the integrated Sachs-Wolfe (ISW) effect~\cite{Sachs:1967er} (see Ref.~\cite{Nishizawa:2014vga} for a recent review) in the framework of the original Hassan-Rosen theory. The ISW effect is the modification of the cosmic microwave background (CMB) spectrum due to the time evolution of gravitational potentials generated by the large-scale structure of the Universe. When CMB photons, traveling between the surface of last scattering and our detectors, enter and then leave a gravitational potential well, they blueshift and redshift, respectively. If the gravitational potential does not evolve with time, which is the case in a flat, matter-dominated universe governed by GR, photons do not lose or gain energy when leaving gravitational wells, and the CMB spectrum is unmodified. If the Universe is not flat or matter-dominated, the gravitational potential does evolve with time, resulting in a net change in the photon energy. Therefore, in GR, the ISW effect is closely linked to dark energy (or curvature). As dark energy starts to dominate and the cosmic expansion starts to accelerate, the potential wells become shallower with time and CMB photons gain a net blueshift.  Since the acceleration of the expansion becomes important at low redshifts, the ISW effect is a late-time phenomenon.\footnote{There also exists a subdominant early-time ISW effect due to the slight impact of the radiation component on the background expansion at the time of decoupling. We disregard this effect in this paper.} Measurements of the ISW effect therefore provide a complementary method to investigate the properties of dark energy. A similar effect is expected if the acceleration is due not to dark energy but rather to large-scale modifications to GR, such as in bigravity, where dark energy is replaced by non-standard gravitational effects. However, since the gravitational potential in these theories behaves differently at late times than in GR, the ISW effect can be different, and measuring its properties can provide another observational tool to distinguish modified gravity from GR and dark energy models. 

The ISW effect is normally not measurable directly from the CMB spectrum~\cite{Huffenberger:2004tv}, since it is subdominant compared to the ordinary Sachs-Wolfe effect caused by gravitational redshift at the time of decoupling. Therefore in order to measure the ISW effect, one has to cross-correlate the CMB temperature fluctuations with other tracers of the gravitational potential, such as the large-scale distribution of galaxies and quasars~\cite{Crittenden:1995ak}. This has motivated several analyses trying to detect the ISW signal using either the angular cross-power spectrum between the CMB temperature anisotropies and large-scale structure, or a similar correlation in pixel space using stacking techniques (see, e.g., Refs.~\cite{Boughn:2003yz,Nolta:2003uy,Fosalba:2003iy,Fosalba:2003ge,Scranton:2003in,Afshordi:2003xu,Huffenberger:2004tv,Vielva:2004zg,Padmanabhan:2004fy,Cabre:2006qm,Pietrobon:2006gh,Rassat:2006kq,McEwen:2007rz,Ho:2008bz,Giannantonio:2008zi,Granett:2008ju,Francis:2009ps,Sawangwit:2009gd,Dupe:2010zs,Goto:2012yc,Schiavon:2012fc,Giannantonio:2012aa,Kovacs:2013rs,Cai:2013ik,Giannantonio:2013uqa,Hernandez-Monteagudo:2013vwa,Ade:2013dsi,Ferraro:2014msa,Aiola:2014cna} and references therein). The significance of detection has been around $3\sigma$ on average, with the strongest detection ($\sim$4.5$\sigma$) reported in Ref.~\cite{Giannantonio:2008zi}. Interestingly, many of these previous studies reported a cross-correlation signal which is systematically larger, by about $1\sigma$--$2\sigma$, than the value predicted in the standard $\Lambda$CDM model (see, e.g., Refs.~\cite{Ho:2008bz,Goto:2012yc,Giannantonio:2008zi,Granett:2008ju,Ade:2013dsi}), with some studies going even further and claiming an observed signal $>3\sigma$ above the predicted standard model value~\cite{Nadathur:2011iu,Flender:2012wu,Hotchkiss:2014lga} using stacking techniques. Note, however, that later studies detected a cross-correlation signal consistent with the prediction of $\Lambda$CDM \cite{Ferraro:2014msa}.

This paper is organized as follows. In Sec.~\ref{sec:HR} we present the Hassan-Rosen theory of bigravity, and in Sec.~\ref{sec:back} we review its predictions for cosmic expansion histories, focusing on the specific case of IBB. The perturbative framework is briefly presented in Sec.~\ref{sec:pert}, and in Sec.~\ref{sec:ISW} we discuss the ISW effect and how it can be approached in massive bigravity. We derive the theoretical ISW contributions to the CMB power spectrum, as well as to the cross-correlation between the CMB anisotropies and large-scale structure. Our numerical results are presented in Sec.~\ref{sec:results}, where they are compared to the observational measurements of the ISW effect. We finally discuss our findings and conclude in Sec.~\ref{sec:conc}.

\section{\label{sec:HR} The theory of massive bigravity}
The Hassan-Rosen theory of ghost-free, massive bigravity has the Lagrangian \cite{Hassan:2011zd}
\begin{align}
\label{eq:HRaction}
\mathcal{L}=&-\frac{M_{g}^{2}}{2}\sqrt{-\det g}R_{g}-\frac{M_{f}^{2}}{2}\sqrt{-\det f}R_{f}\nonumber \\
&+m^{4}\sqrt{-\det g}\sum_{n=0}^{4}\beta_{n}e_{n}\left(\sqrt{g^{-1}f}\right)+\sqrt{-\det g}\mathcal{L}_\mathrm{m},
\end{align}
where $R_{g}$ and $R_{f}$ are the Ricci scalars for the metrics $g_{\mu\nu}$ and $f_{\mu\nu}$, respectively. The functions $e_{n}$ are the elementary symmetric polynomials of the eigenvalues of the matrix $\sqrt{g^{-1}f}$, where the matrix square root is defined such that $\sqrt{g^{-1}f}\sqrt{g^{-1}f}=g^{\mu\nu}f_{\mu\nu}$. The forms of these polynomials are given in, e.g., Ref.~\cite{Hassan:2011zd}. The quantities $\beta_n$ $(n=0,1,2,3,4)$ are free parameters of the theory to be constrained by observations. As pointed out earlier, in this paper we consider the so-called singly-coupled bigravity theory where the matter Lagrangian $\mathcal{L_\mathrm{m}}$ couples only to $g_{\mu\nu}$.

The generalized Einstein equations for the two metrics are
\begin{align}
G^g_{\mu\nu}+m^{2}\sum_{n=0}^{3}\left(-1\right)^{n}\beta_{n}g_{\mu\lambda}Y_{\left(n\right)\nu}^{\lambda}\left(\sqrt{g^{-1}f}\right)&=\frac{1}{M_g^2}T_{\mu\nu},\\
G^f_{\mu\nu}+m^2\sum_{n=0}^{3}\left(-1\right)^{n}\beta_{4-n}f_{\mu\lambda}Y_{\left(n\right)\nu}^{\lambda}\left(\sqrt{f^{-1}g}\right)&=0,
\end{align}
where the Einstein tensors corresponding to $g_{\mu\nu}$ and $f_{\mu\nu}$ are defined as usual, $G^g_{\mu\nu}=R^g_{\mu\nu}-\frac 12R_gg_{\mu\nu}$ with $R^g_{\mu\nu}$ the $g$-metric Ricci tensor and similarly for $G^f_{\mu\nu}$, and the matrices $Y_{\left(n\right)}$ are functions of $\sqrt{g^{-1}f}$ which are defined, e.g., in Ref.~\cite{Hassan:2011zd}. $T_{\mu\nu}$ is the stress-energy tensor of matter defined in the usual way with respect to the physical metric, $g_{\mu\nu}$. Here we have performed a rescaling of the parameters and fields so that $M_f^2$ becomes a redundant parameter.\footnote{See, however, Ref.~\cite{Akrami:2015qga}, which appeared during the completion of this work, for caveats associated with this rescaling.} This is possible when we couple matter only to $g_{\mu\nu}$. Finally, by combining the Bianchi identity with the field equations and the conservation of the stress-energy tensor, we arrive at the Bianchi constraint on the interaction (mass) terms,
\be
 \nabla_g^{\mu}\frac{m^2}{2}\sum_{n=0}^3\left(-1\right)^{n}\beta_ng_{\mu\lambda}Y^{\lambda}_{(n)\nu}\left(\sqrt{g^{-1}f}\right) = 0, \label{eq:Bianchig}
\ee
where $\nabla_g$ is the covariant derivative operator with respect to $g_{\mu\nu}$.

\section{Cosmological background}
\label{sec:back}

Following the literature on the cosmology of massive bigravity, and assuming the cosmological principles of homogeneity and isotropy, we assume here that the Universe at the background level can be described by FLRW metrics for both $g_{\mu\nu}$ and $f_{\mu\nu}$.\footnote{See Ref.~\cite{Nersisyan:2015oha} and references therein for other possible metrics in bimetric cosmology.} For reasons of brevity, we do not present the details of how to derive the background solutions, and instead refer the reader to, e.g., Refs~\cite{Volkov:2011an,Comelli:2011zm,vonStrauss:2011mq,Akrami:2012vf,Akrami:2013pna,Konnig:2013gxa}.

The background has three free metric functions: the $g$-metric scale factor $a_g$, the $f$-metric lapse $N_f$, and the $f$-metric scale factor $a_f$. The $g$-metric lapse $N_g$ can be chosen arbitrarily through a choice of time coordinate; we work with conformal time in this paper, i.e., we set $N_g=a_g$. 
The matter sector is described by the energy density $\rho$ and the pressure $p$, which satisfy the usual continuity equation with respect to the metric $g_{\mu\nu}$,
\be\label{eq:cont}
\dot{\rho}+3\frac{\dot{a}_g}{a_g}\left(\rho+p\right)=0.
\ee
The Bianchi constraint (\ref{eq:Bianchig}) yields two possible solution branches. The first branch places an algebraic constraint on the quantity $r\equiv a_f/a_g$, which becomes a constant with a value that depends on the parameters $\beta_1$, $\beta_2$ and $\beta_3$. The second branch places a dynamical constraint on the $f$-metric lapse: $N_f=a_g\dot{a}_f/\dot{a}_g$. Focusing on the second (dynamical) branch, which gives interesting non-GR solutions, and defining
\begin{align}
\label{eq:B0}
B_{0}(r)&\equiv\beta_{0}+3\beta_{1}r+3\beta_{2}r^2+\beta_{3}r^3, \\
\label{eq:B1}
B_{1}(r)&\equiv\beta_{1}r^{-3}+3\beta_{2}r^{-2}+3\beta_{3}r^{-1}+\beta_{4},
\end{align}
the field equations become 
\begin{align}
\label{eq:friedg}
3H_{g}^2 &= \frac{\rho}{M_g^2} + m^2 B_0, \\
\label{eq:poly}
\frac{\rho}{M_g^2}&=m^2\left(B_1r^2-B_0\right),
\end{align}
where $H_{g} \equiv \dot a_{g}/a_{g}^2$ is the Hubble factor for the physical metric $g_{\mu\nu}$. The first equation is the standard Friedmann equation modified by a mass term that depends on $r$, while the second equation determines $r$ algebraically.

As mentioned in Sec.~\ref{sec:intro}, in this paper we focus on the IBB model where only $\beta_1$ and $\beta_4$ are non-vanishing. This model contains no explicit cosmological constant term, has stable linear perturbations, and is a viable alternative to $\Lambda$CDM in terms of background expansion and growth of structure~\cite{Konnig:2014xva}. In IBB, $r$ starts off from a large value at early times and decreases with time to a finite value in the far future. The best-fit parameter values for this model from the comparison of its theoretical predictions to the cosmic expansion history, as well as the linear growth of structure, are $\beta_1=0.48$ and $\beta_4=0.94$~\cite{Solomon:2014dua,Konnig:2014xva},\footnote{Note, however, that background observations can only pick out a curve in the $\beta_1$-$\beta_4$ space, and while perturbations can in principle break this degeneracy, present data do so poorly, cf. Fig. 3 in Ref.~\cite{Konnig:2014xva}. Therefore while the parameters quoted here fit the best, a range of other parameters along this best-fit curve fits nearly as well.} where the parameters are measured in units of $H_{0}^2$ with $H_0\equiv H_{g,0}$ the Hubble rate today, and the mass parameter $m^2$ is absorbed into the parameters $\beta_n$. These are the parameter values which we use in this paper when comparing the ISW predictions to observations.

\section{Cosmological perturbations}
\label{sec:pert}

Cosmological perturbations in massive bigravity have been derived and studied in several previous works. They were first presented in Ref.~\cite{Comelli:2012db}, and the scalar perturbations were then further studied in Ref.~\cite{Khosravi:2012rk} in the superhorizon limit and in connection with primordial perturbations. The perturbative equations were rederived in Ref.~\cite{Berg:2012kn} and studied in the future de Sitter regime. In Ref.~\cite{Konnig:2014dna}, the linear growth of cosmological structure in the quasi-static limit and for subhorizon modes was studied for a minimal bigravity model, where only the parameter $\beta_1$ is non-vanishing. It was shown that this minimal model, despite possessing a viable expansion history, is unstable at early times on small scales, confirming the instabilities found previously in Ref.~\cite{Comelli:2012db}. An extensive analysis of the theory for scalar perturbations and for various combinations of the parameters was then performed in Ref.~\cite{Solomon:2014dua}, the quasi-static limit for subhorizon scales was studied, several modified gravity parameters were calculated, and deviations from GR predictions measurable by future large-scale structure experiments were presented. Ref.~\cite{Konnig:2014xva} then studied the problem of instabilities in further detail, and identified specific submodels and parameter combinations which avoid the instabilities; this introduced the IBB model that we consider in this paper. The problem of gradient instabilities has been further studied in Ref.~\cite{DeFelice:2014nja}. Finally, Ref.~\cite{Lagos:2014lca} presented a comprehensive analysis of the scalar, vector, and tensor cosmological perturbations, confirming the results found in the previous studies. Cosmological perturbations have also been studied in some extensions of the theory; see, e.g., Ref.~\cite{Comelli:2014bqa} for the case where the $f$ metric is coupled to a second matter sector.

In order to obtain the perturbation equations used in our calculation of the ISW effect for bigravity, as will be discussed in the next section, we follow Refs.~\cite{Solomon:2014dua,Konnig:2014xva,Lagos:2014lca} and use the following ansatz for the perturbative metrics:
\begin{align}
ds_{g}^{2}=&-a_g^{2}\left(1+E_{g}\right)d\eta^{2}+2a_g^{2}\partial_{i}F_{g}d\eta dx^{i} \nonumber \\
&+a_g^{2}\left[\left(1+A_{g}\right)\delta_{ij}+\partial_{i}\partial_{j}B_{g}\right]dx^{i}dx^{j},
\end{align}
\begin{align}
ds_{f}^{2}=&-N_f^{2}\left(1+E_{f}\right)d\eta^{2}+2N_f a_f\partial_{i}F_{f}d\eta dx^{i} \nonumber \\
&+a_f^{2}\left[\left(1+A_{f}\right)\delta_{ij}+\partial_{i}\partial_{j}B_{f}\right]dx^{i}dx^{j}.
\end{align}
The equations of motion for the perturbation variables and their numerical solutions have been given in the same references (see also Refs.~\cite{Comelli:2012db,Khosravi:2012rk,Berg:2012kn,Comelli:2014bqa,DeFelice:2014nja}). Since the equations are rather lengthy, we refer the reader to those papers for details and here we only present some key definitions and assumptions used in our ISW analysis.

Forming the gauge-invariant variables~\cite{Konnig:2014dna,Konnig:2014xva}
\begin{align}
\Phi&\equiv\frac{1}{2}E_{g}-\frac{1}{a_{g}}\partial_{\eta}\left[a_g\left(\frac{1}{2}\dot{B}_{g}-F_{g}\right)\right],\\
\Psi&\equiv-\frac{1}{2}A_{g}+\frac{\dot a_{g}}{a_{g}}\left(\frac{1}{2}\dot{B}_{g}-F_{g}\right),\\
\delta &\equiv \frac{\delta\rho_m}{\rho_m}+ \frac{3\dot a_{g}}{a_{g}}\left(\frac{1}{2}\dot{B}_g-F_g\right),
\end{align}
where $\rho_m$ is the background energy density of matter and $\delta\rho_m$ is its perturbation, the equations of motion can be written as two second-order differential equations for the gravitational potentials $\Phi$ and $\Psi$. Having solved these numerically (in the gauge $F_g=F_f=0$), we can obtain the time evolution of the gauge-invariant density contrast of the matter perturbations.

$\Phi$, $\Psi$ and $\delta$ are the quantities needed when computing the ISW effect. We again emphasize that the model we study in this paper is IBB, therefore the variables and the corresponding equations of motion we compute are the ones specific to the IBB model.

\begin{figure}
\begin{centering}
\includegraphics[scale=0.68]{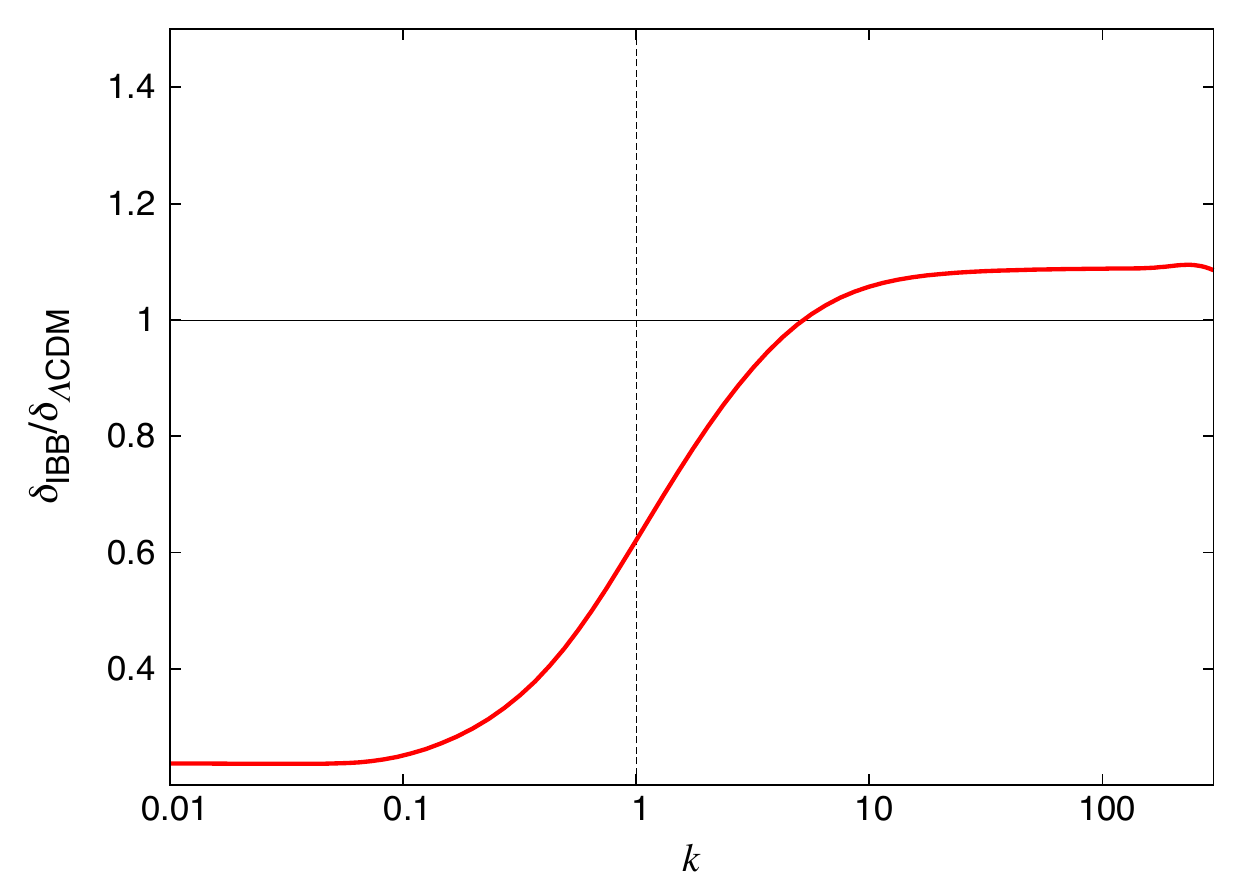}
\par\end{centering}
\caption{\label{fig:delta}Ratio of the gauge-invariant density contrast in IBB today to the one in $\Lambda$CDM, as a function of scale, computed using the same initial conditions for the gravitational potentials at the time of decoupling. Scales $k$ are in units where $H_0=1$.}
\end{figure}

Before moving on to the discussion of the ISW effect in bigravity compared to that in $\Lambda$CDM, it is illustrative to see how the density contrast of matter perturbations differs between the two models on different scales. We therefore plot in Fig.~\ref{fig:delta} the ratio of the density contrasts today computed in $\Lambda$CDM (with $\Omega_m=0.3$) and IBB (with $\beta_1=0.48$ and $\beta_4=0.94$). We have used the same initial conditions in both models for the gravitational potentials at the time of decoupling. The figure shows that it is only for modes close to or larger than the horizon scale today that the predictions of the two models differ significantly, with IBB predicting much less superhorizon structure than $\Lambda$CDM. As we will see in the following sections, the amplitude of the ISW effect is large only at low multipoles ($\ell<100$); the multipoles $2<\ell<10$ receive contributions from the modes $1<k<100$, where the scales $k$ are measured in units of $H_0$.

\section{The ISW effect}
\label{sec:ISW}

The ISW effect is due to the time evolution of gravitational potentials $\Phi$ and $\Psi$ between the epoch of recombination (last scattering surface) and today, generating the following secondary anisotropies on the CMB sky:
\be
\left(\frac{\delta T}{T}\right)_\mathrm{ISW}=\intop_{\eta_{r}}^{\eta_{0}}\left(\dot{\Psi}+\dot{\Phi}\right)d\eta.
\ee 

As mentioned in Sec.~\ref{sec:intro}, the ISW effect is measurable through the cross-correlation between the CMB anisotropies and the large-scale structure. Therefore, in order to compare the predictions of bigravity to observations, we need to compute this cross-correlation. The standard quantity to work with is the cross-correlation angular power spectrum, $C^{Tg}_\ell$, since most of the ISW analyses are done in harmonic space. Following Ref.~\cite{Ferraro:2014msa}, $C^{Tg}_\ell$ is given by
\be\label{eq:CTg}
C^{Tg}_\ell=C^{\dot{\Phi}g}_\ell=4\pi\int\frac{dk}{k}\Delta_m^2(k)K^{\dot{\Phi}}_\ell(k)K^g_\ell(k).
\ee
This form for $C^{Tg}_\ell$ assumes that there are no other contributions to the correlation between the CMB temperature anisotropies and large-scale structure at low redshifts. Here, $K^{\dot{\Phi}}_\ell$ and $K^{g}_\ell$ are weight functions given by
\begin{align}
K_{\ell}^{g}(k)&=\int dzb(z)\frac{dN}{dz}\frac{\delta(k,z)}{\delta(k,0)}j_{\ell}\left[k\chi(z)\right],\label{eq:Kg}\\
K_{\ell}^{\dot{\Phi}}(k)&=\int dz\frac{d}{dz}\left[\frac{\Phi+\Psi}{\delta(k,0)}\right]j_{\ell}\left[k\chi(z)\right],\label{eq:KPhid}
\end{align}
where $b(z)$ is the bias factor which relates tracers to the underlying mass distribution, $dN/dz$ is the redshift distribution of the density tracers normalized so that $\int dz \frac{dN}{dz}=1$, and $j_\ell$ are the spherical Bessel functions with arguments depending on $\chi(z)=\eta_0-\eta(z)$, where $\eta$ is conformal time and $z=1/a-1$ is the redshift. $\Delta_m^2(k)$ in Eq. (\ref{eq:CTg}) is the power spectrum of the matter density fluctuations today.

In GR, one usually introduces a $k$-independent growth function $D(z)$, normalized such that $D(z=0)=1$. The density contrast is then given by $\delta(k,z)=D(z)\delta(k,0)$. In addition, due to the absence of anisotropic stress in GR, the potentials satisfy $\Phi=\Psi$, and for subhorizon modes one has
\be\label{eq:potdelta}
\Phi=\frac{3}{2}\frac{\Omega_{m}}{a}\left(\frac{H_{0}}{k}\right)^{2}\delta(k,z),
\ee
where $\Omega_{m}$ is the matter density parameter at present time. Plugging these relations and equations in Eqs. (\ref{eq:Kg}) and (\ref{eq:KPhid}), one recovers the known expressions for the weight functions in GR.

In massive bigravity, $\Phi$ is no longer equal to $\Psi$ at all times, and furthermore, the time evolutions of the density contrast and potentials depend on $k$. It is possible to normalize the weight functions with respect to either the density contrast today, or the potentials at the time of decoupling. We explore both options below. For the power spectrum, we use the standard form
\be\label{eq:deltam}
\Delta_{m}^{2}=\delta_{H}^{2}\left(\frac{k}{H_{0}}\right)^{n+3}T^{2}(k),
\ee
where $\delta_H$ is the amplitude of the density contrast at the horizon scale today, $n$ is the spectral index, and $T(k)$ is the transfer function. For the transfer function we use the Bardeen-Bond-Kaiser-Szalay (BBKS) approximation \cite{Bardeen:1985tr}. Normalizing the density contrast with respect to the potentials at the time of decoupling, one can use Eq. (\ref{eq:potdelta}) to infer the spectrum of $\Phi$ at that time. One then replaces $\Delta_m^2$ with $\Delta_\Phi^2$ in Eq. (\ref{eq:CTg}), and normalizes the weight functions with respect to the potentials at the time of decoupling instead of the density contrast today. In GR, both approaches obviously give the same results, but in massive bigravity $C^{Tg}_\ell$ varies slightly depending on the choice of the normalization scheme.

Besides the cross-correlation spectrum $C^{Tg}_\ell$, one can also compute the ISW auto-correlation spectrum $C_{\ell}^\mathrm{ISW}$ directly, although, as mentioned earlier, this quantity cannot be used to detect the ISW effect from observations. $C_{\ell}^\mathrm{ISW}$ is given by
\be\label{eq:Cpdotpdot}
C_{\ell}^\mathrm{ISW}\equiv C_{\ell}^{\dot{\Phi}\dot{\Phi}}=4\pi\int\frac{dk}{k}\Delta_{\Phi}^{2}(k)K_{\ell}^{\dot{\Phi}}(k)K_{\ell}^{\dot{\Phi}}(k).
\ee

In order to calculate $C^{Tg}_\ell$ and $C_{\ell}^\mathrm{ISW}$ using Eqs. (\ref{eq:CTg}) and (\ref{eq:Cpdotpdot}), we need to compute the time evolution of the gravitational potentials $\Phi$ and $\Psi$. As discussed in Sec. \ref{sec:pert}, we write the equations of motion for the potentials as two second-order differential equations (see Refs.~\cite{Konnig:2014xva,Lagos:2014lca} for details) and solve them numerically. We set the initial conditions at the time of decoupling and then evolve the fields to the present time. We additionally assume $\Phi=\Psi$ at the time of decoupling. The numerical analysis is performed using Mathematica.

\section{Comparison to data}
\label{sec:results}

A recent analysis of the ISW cross-correlation has been performed in Ref.~\cite{Ferraro:2014msa} using the 9-year {\it Wilkinson Microwave Anisotropy Probe} (WMAP) CMB temperature data~\cite{Bennett:2012zja}, as well as a sample of galaxies and quasars provided by the {\it Wide-field Infrared Survey Explorer} (WISE)~\cite{Wright:2010qw}. WISE scanned the entire sky in four frequency bands and provided a catalog which contains more than 500 million sources, including galaxies with redshifts up to $z\sim1$ and a median redshift of 0.3~\cite{Yan:2012yk}; the dataset is therefore much deeper than the ones provided by other experiments at similar frequencies. It is argued in Ref.~\cite{Ferraro:2014msa} that the large number of sources WISE detected, together with the large area of the survey and its redshift distribution, makes the catalog one of the best for measuring the ISW effect through its cross-correlation with the CMB temperature anisotropies. WISE had been used previously for the study of the ISW effect (see, e.g., Refs.~\cite{Goto:2012yc,Kovacs:2013rs}), but Ref.~\cite{Ferraro:2014msa} uses a larger sample by applying less-conservative cuts to the dataset and a higher median redshift. For these reasons, we use the results of Ref.~\cite{Ferraro:2014msa} in order to compare the ISW predictions of IBB to the observations. 

Before moving on to the discussion of our numerical analysis and the results, we note that care must be taken regarding the bias factor $b(z)$ used in Eq. (\ref{eq:Kg}). Following Ref.~\cite{Ferraro:2014msa}, we use the measured cross-power of galaxies and AGN from the WISE survey with the CMB temperature data. Galaxies and quasars trace the gravitational potential up to the bias factor. It is therefore crucial to properly measure the bias when comparing the predictions of the ISW effect in a theoretical model to the observational measurements. It is argued in Ref.~\cite{Ferraro:2014msa} that the best way to measure the bias is through cross-correlating lensing potential maps with the tracer field. The reason is that gravitational lensing is directly sourced by the gravitational potential, and therefore the measurement of the bias is more robust and less influenced by systematic errors which are present in measurements using the auto-correlation spectrum; these errors can lead to an incorrect estimation of the amplitude of the ISW effect. The authors therefore used weak lensing of the CMB to measure the effective bias. The data used for this procedure are the Planck lensing potential~\cite{Ade:2013tyw} as well as WISE maps, and the cross-correlation signal is measured for $100<\ell<400$ (see Refs.~\cite{Ade:2013tyw,Geach:2013zwa} for earlier studies of the correlation between WISE and the Planck lensing potential). Ref.~\cite{Ferraro:2014msa} assumes two models for the galaxy bias factor as a function of redshift; for a redshift-independent bias, $b^G$, their estimated value is $b^G=1.41\pm0.15$, and for a bias linearly dependent on redshift, $b^G(z)=b_0^G(1+z)$, they find $b_0^G=0.98\pm0.10$. For AGN, the bias model is $b^A(z)=b_0^A[0.53+0.289(1+z)^2]$ with $b_0^A=1.26\pm0.23$. We use the same estimated bias factors in the analysis of IBB.

With the measured values of the bias factor discussed above, the ISW effect has been detected in Ref.~\cite{Ferraro:2014msa} at the level of $3\sigma$, with a cross-correlation amplitude, normalized to the expected correlation in a $\Lambda$CDM universe, of $\mathcal{A}=1.24\pm 0.47$ for the constant bias and $\mathcal{A}=1.54\pm 0.59$ for the linear bias. In both cases 50 million galaxies (with a redshift distribution peak at $z\sim 0.3$) have been used. The cross-correlation amplitude using one million AGN (with a redshift distribution peak at $z\sim 1.1$) has been found to be $\mathcal{A}=0.88\pm 0.74$.

We now apply the method of calculating the ISW spectra discussed in Sec. \ref{sec:ISW} to our bigravity model, IBB. This is similar to the method employed in Ref.~\cite{Ferraro:2014msa} for $\Lambda$CDM. We use the same bias factors for galaxies and AGN as in Ref.~\cite{Ferraro:2014msa}, and the redshift distributions $dN/dz$ for the galaxies and AGN are taken from Refs.~\cite{Yan:2012yk,Geach:2013zwa}.

\begin{figure}
\begin{centering}
\includegraphics[scale=0.68]{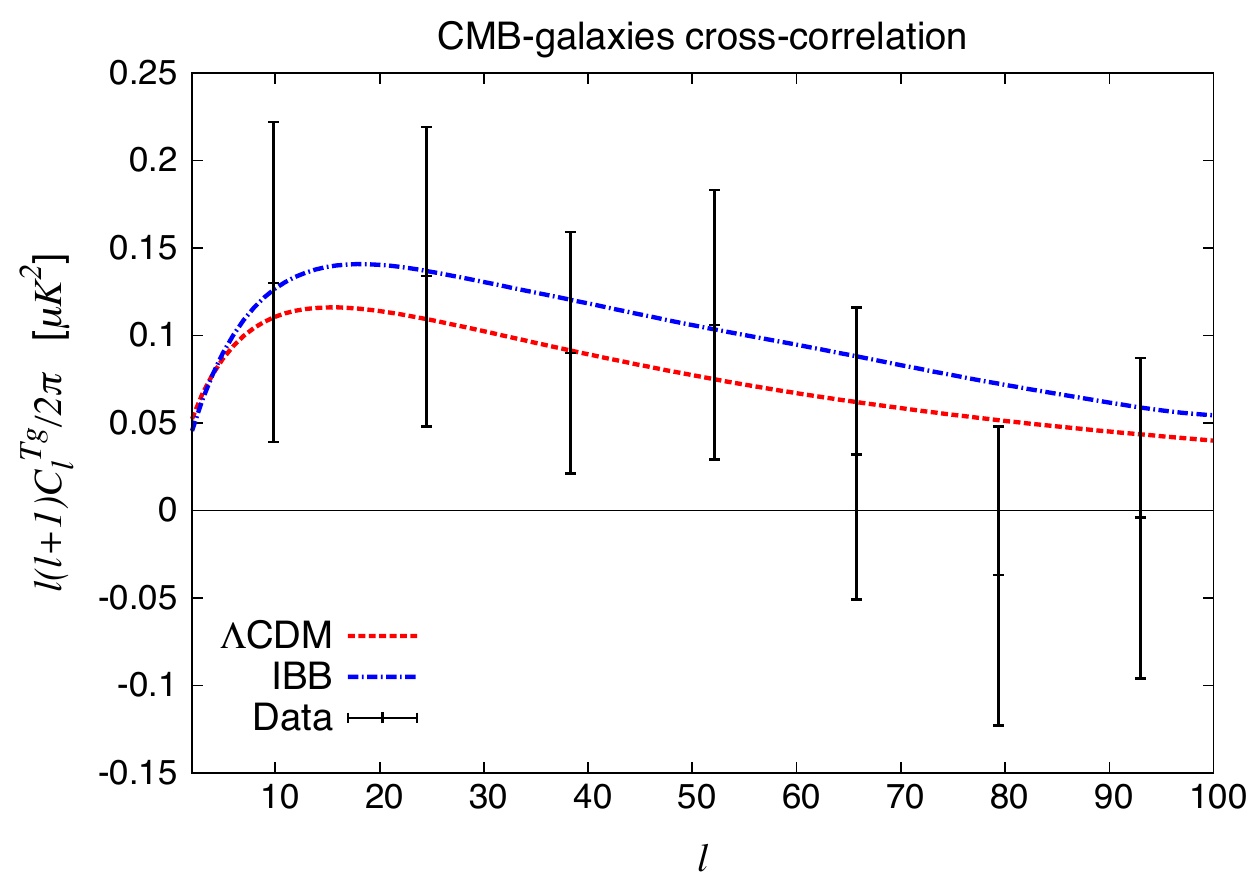}
\includegraphics[scale=0.68]{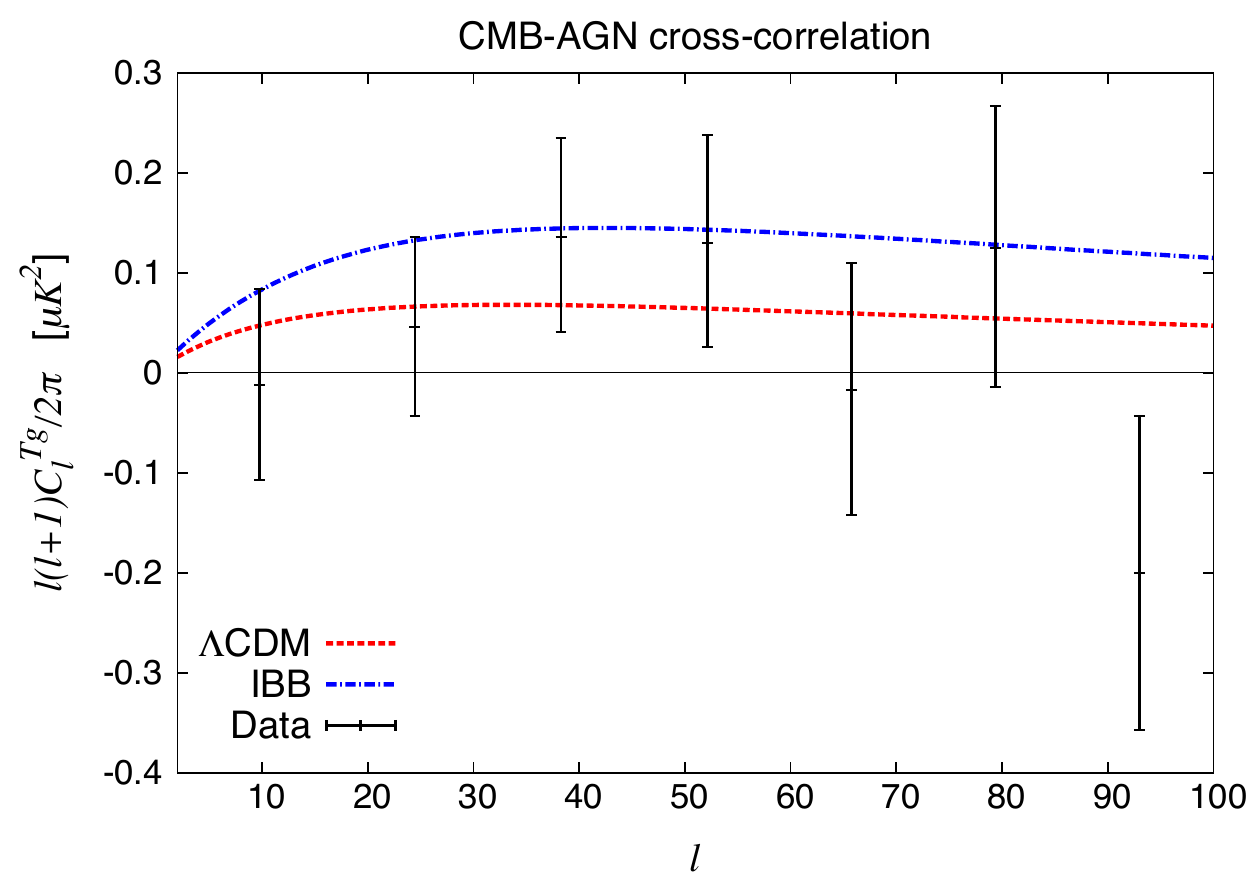}
\par\end{centering}
\caption{\label{fig:CTg} {\bf Top panel:} Theoretical cross-correlation power spectrum $C^{Tg}_\ell$ between the CMB temperature anisotropies and galaxies computed for $\Lambda$CDM and IBB using the density contrast today to normalize the weight functions, as well as the data points measured using the WISE galaxy catalog in Ref.~\cite{Ferraro:2014msa}. {\bf Bottom panel:} The same for the cross-correlation spectrum between the CMB temperature anisotropies and AGN.}
\end{figure}

In Fig.~\ref{fig:CTg} we plot the theoretical cross-correlation power spectrum $C^{Tg}_\ell$ between the CMB temperature anisotropies and galaxies/AGN computed for both $\Lambda$CDM ($\Omega_m=0.3$) and IBB ($\beta_1=0.48$ and $\beta_4=0.94$), together with the measured values using the WISE galaxy and AGN catalogs and the corresponding uncertainties as reported in Ref.~\cite{Ferraro:2014msa}. Here we have used the density contrast today to normalize the weight functions (\ref{eq:Kg}) and (\ref{eq:KPhid}), assumed $\delta_H=7.5\times10^{-5}$ and $n=1$ [see Eq. (\ref{eq:deltam})], and used the constant bias for galaxies. The figure shows that IBB predicts a higher correlation (by roughly a factor of 1.5) compared to $\Lambda$CDM, but the amplitudes are still consistent with the data. A similar conclusion can be made when the weight functions are normalized using the gravitational potentials at the time of decoupling or when the linearly evolving bias for the galaxies is used. Note that our theoretical curves for $\Lambda$CDM are slightly different from the ones presented in Ref.~\cite{Ferraro:2014msa}, which is due to the difference in the choice of transfer function. In principle, one could perform the same calculation by using parameter values closer to the ones from WMAP9 or Planck and by choosing a different transfer function, but this would only change the theoretical curves slightly. Our main objective in this paper is comparing $\Lambda$CDM and IBB in terms of the ISW predictions and what is important in our analysis is to calculate the theoretical curves using the same initial conditions for the two models, and infer the relative shift in the ISW amplitudes.

\begin{figure}[t]
\begin{centering}
\includegraphics[scale=0.68]{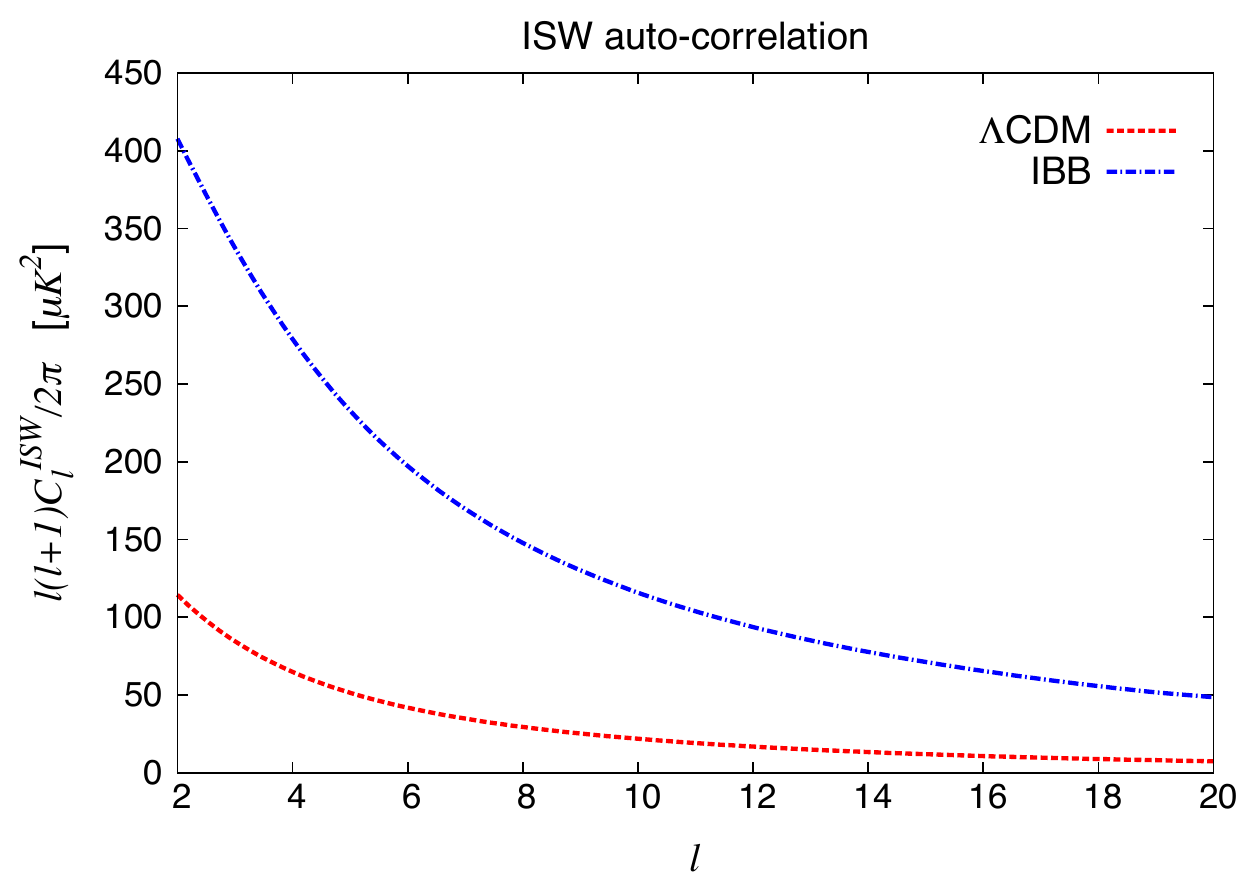}
\par\end{centering}
\caption{\label{fig:Cpdotpdot}Theoretical ISW auto-correlation power spectrum $C_{\ell}^\mathrm{ISW}$ for both $\Lambda$CDM and IBB.}
\end{figure}

It is also interesting to see the theoretical ISW auto-correlation power spectrum $C_{\ell}^\mathrm{ISW}$ [defined in Eq. (\ref{eq:Cpdotpdot})] for both $\Lambda$CDM and IBB; this is shown in Fig.~\ref{fig:Cpdotpdot}. There are no data points on that plot, since, as we mentioned earlier, the signal is too small to be detected directly in the CMB spectrum. The figure once again shows that the ISW effect in IBB is larger than that in $\Lambda$CDM. It is however still subdominant compared to the primary CMB signal and is consistent with the data given the large uncertainties on large scales due to cosmic variance.

Before we end this section, let us present another way of comparing the ISW amplitudes in $\Lambda$CDM and IBB, this time in terms of how well the amplitudes can be measured from the same sets of observations. This is based on the expected values of signal-to-noise ratio in the two models. Following a Fisher matrix analysis presented in Ref.~\cite{Ferraro:2014msa}, the signal-to-noise ratio is:

\begin{align}
\left(\frac{S}{N}\right)^2 &\approx f_\mathrm{sky} \sum_\ell (2\ell+1)\frac{[C_\ell^{Tg}]^2}{C_\ell^{TT}C_\ell^{gg}+[C_\ell^{Tg}]^2}\nonumber \\
&\approx f_\mathrm{sky} \sum_\ell (2\ell+1)\frac{[C_\ell^{Tg}]^2}{C_\ell^{TT}C_\ell^{gg}},
\end{align}
where $f_\mathrm{sky}$ is the sky coverage fraction, $C_\ell^{Tg}$ is again the cross-correlation power spectrum between the CMB temperature anisotropies and large-scale structure (galaxies or AGN), and $C_\ell^{TT}$ and $C_\ell^{gg}$ are the auto-correlation power spectra for the CMB and large-scale structure, respectively. The first thing one can infer from this expression for the signal-to-noise is that almost all of the contributions to the ISW effect are from low multipoles and low redshifts: they are negligible for $\ell>100$ and $z>1.5$~\cite{Afshordi:2004kz}. In Ref.~\cite{Ferraro:2014msa} a theoretical maximum value for the signal-to-noise ratio has been evaluated to be $7.6\sqrt{f_\mathrm{sky}}$ for $\Lambda$CDM; this is the maximum signal-to-noise one can achieve taking into account the uncertainties coming from cosmic variance. 

Repeating the same calculation of the signal-to-noise ratio for IBB, we can compare that value to the one for $\Lambda$CDM. We find that the ratio of the signal-to-noise as computed for IBB and $\Lambda$CDM, i.e. $(S/N)_\mathrm{IBB}/(S/N)_{\Lambda\mathrm{CDM}}$, is $\sim$1.2 for the galaxies, and $\sim$1.6 for the AGN. These ratios are closely related to ratio of the cross-correlation amplitudes $\mathcal A_\mathrm{IBB}$ and $\mathcal A_{\Lambda\mathrm{CDM}}$ for IBB and $\Lambda$CDM, respectively, and show that the difference between the models is well within the observational uncertainties.

\section{Discussion and conclusions}
\label{sec:conc}

In this paper we have studied the integrated Sachs-Wolfe (ISW) effect in the Hassan-Rosen theory of bimetric massive gravity. The ISW effect is sensitive to the time evolution of the gravitational potentials sourced by large-scale structure and is nonzero at late times when the cosmic expansion is accelerating. It therefore offers a complementary tool for probing the physics governing the late-time evolution of the Universe and distinguishing between possible mechanisms generating the cosmic acceleration; this includes both dark energy and modified gravity theories. The ISW effect contributes to large scales (low multipoles) of the CMB temperature auto-correlation power spectrum as well as the cross-correlation power spectrum between the CMB and tracers of large-scale structure such as galaxies and quasars. The contribution to the auto-correlation is subdominant compared to the primordial contributions and is therefore too small to be detected observationally. The cross-correlation contributions are however large enough and have been detected by various studies of the effect.

We have focused on the infinite-branch bigravity (IBB) model, which has viable cosmological solutions in terms of both expansion histories and linear growth of structure. The model is also free from gradient instabilities and can explain the late-time expansion of the Universe without resorting to an explicit cosmological constant, and therefore offers an interesting alternative to the standard cosmological $\Lambda$CDM model.

The background and perturbative equations in bigravity in general and IBB in particular were reviewed, emphasizing the quantities and equations required for our ISW analysis. We then presented our method for studying the ISW effect in bigravity in terms of the cross-correlation between the CMB temperature anisotropies and large-scale structure, which we employed to calculate the theoretical predictions for a best-fit IBB model and compare them to both the predictions in $\Lambda$CDM and observational measurements of the effect.

The data we have used in our analysis are taken from Ref.~\cite{Ferraro:2014msa}, where the 9-year WMAP CMB temperature maps as well as the datasets provided by the WISE survey were analyzed to measure the ISW effect. We used the measurements performed in Ref.~\cite{Ferraro:2014msa} for the WISE catalogs of both galaxies and AGN. We also used the bias factors provided by Ref.~\cite{Ferraro:2014msa}.

We applied our method of computing the ISW auto- and cross-correlation amplitudes to the IBB model, and compared our results to those of $\Lambda$CDM and to the observational measurements. We found that IBB predicts higher amplitudes compared to $\Lambda$CDM for the ISW temperature auto-correlation, as well as the temperature-galaxy and temperature-AGN cross-correlations. The amplitudes are however consistent with the measurements from WISE, and IBB is therefore not ruled out by existing data.

In order to calculate the cross-correlation power spectra, we used two different methods of normalizing the weight functions appearing in the expression for the spectra: the normalization is done with respect to either the gravitational potentials at the time of decoupling or the matter density contrast today. The theoretical amplitudes for IBB depend weakly on the choice of the normalization scheme, and are in both cases consistent with observational measurements. We emphasize that ultimately a more careful analysis of the evolution of the perturbations in a multi-component framework is needed in order to set up the correct normalization scheme. This is beyond the scope of the present paper and we leave it for future work.

\acknowledgments

We are grateful to Luca Amendola, Tomi S. Koivisto, Frank K\"onnig, and the authors of Ref. \cite{Ferraro:2014msa} for useful discussions. Y.A. acknowledges support from DFG through the project TRR33 ``The Dark Universe.'' E.M. acknowledges support for this study from the Swedish Research Council. The work of M.R. has been supported by the DFG cluster of excellence ``Origin and Structure of the Universe.'' A.R.S. is grateful for support from the Cambridge Philosophical Society and the STFC.
\bibliography{bibliography}

\end{document}